\documentclass[twocolumn,showpacs,amsmath,amssymb,10pt]{revtex4}
\usepackage{amsfonts}
\usepackage{bm}

\newcommand{\ot}{{\,\otimes\,}}
\newcommand{{\Cd}}{{\mathbb{C}^d}}

\def\<{\langle}
\def\>{\rangle}
\newtheorem{theorem}{Theorem}

\newtheorem{remark}{Remark}
\newtheorem{corollary}{Corollary}

\begin{document}
\title{\textbf{Witnessing quantum discord in $2 \times N$ systems}}
\author{Bogna Bylicka and Dariusz Chru\'sci\'nski\thanks{email:
darch@phys.uni.torun.pl} }
\affiliation{Institute of Physics, Nicolaus Copernicus University,\\
Grudzi\c{a}dzka 5/7, 87--100 Toru\'n, Poland}

\begin{abstract}

Bipartite states with vanishing quantum discord are necessarily
separable and hence positive partial transpose (PPT). We show
$2\times N$ states satisfy additional property:  the positivity of
their partial transposition is recognized with respect to the
canonical factorization of the original density operator. We call
such states SPPT (for strong PPT). Therefore,  we provide a natural
witness for a quantum discord: if a $2\times N$ state is not SPPT it
must contain nonclassical correlations measured by quantum discord.
It is an analog of the celebrated Peres-Horodecki criterion: if a
state is not PPT it must be entangled.

\end{abstract}

\maketitle

\section{Introduction}


Quantum entanglement is one of the most remarkable features of
quantum mechanics and it leads to powerful applications like quantum
cryptography, dense coding and quantum computing \cite{QIT,HHHH}.
However, a quantum state of a composed system may contain other
types of nonclassical correlation even if it is separable (not
entangled). For a recent `catalogue' of nonclassical correlations
see \cite{Paterek}. The most popular measure of such correlations --
quantum discord -- was introduced by Ollivier and Zurek and
independently by Henderson and Vedral \cite{Zurek, Vedral}. Hence,
quantum discord captures the nonclassical correlations, more general
than entanglement, that can exist between parts of a quantum system
even if the corresponding quantum entanglement does vanish.

Quantum discord has received much attention in studies involving
thermodynamics and correlations \cite{T1,T2},  complete positivity
of reduced quantum dynamics \cite{CP1,CP2} and broadcasting of
quantum states \cite{B1,B2}. It was shown that quantum discord might
be responsible for the quantum computational efficiency of some
quantum computation tasks \cite{Q1,Q2,Q3}. Recently, both Markovian
and non-Markovian dynamics of discord was analyzed \cite{M1,M2}.
Interestingly, contrary to quantum entanglement, Markovian evolution
can never lead to a sudden death of discord. Hence, a generic
quantum state may keep quantum discord forever. Quantum discord was
analytically computed for a class of 2-qubit state
\cite{Lu,X-recent}. Finally, it was recently generalized for systems
with continuous variables to study the correlations in Gaussian
states \cite{G1,G2}. Interestingly, it was shown \cite{Acin} that a
set of states with vanishing discord has vanishing volume in the set
of all states. Actually, this result holds true for any Hilbert
space dimension. It shows that a generic state of composed quantum
system does contain nonclassical correlation.

In the present paper we analyze a class of states of $2 \times N$
systems. Such `qubit-quNit' systems play important role in quantum
information and were intensively analyzed \cite{2N}. Note, that a
state with vanishing quantum discord -- so called classical-quantum
state -- is necessarily separable and hence PPT (positive partial
transpose). Recently, we introduced \cite{SPPT} a subclass of PPT
states -- so called SPPT (strong positive partial transpose). These
are states where the PPT property is guarantied by the canonical
construction based on certain decomposition of the density operator
(see below). It was conjectured that all SPPT states are separable.
Now, we prove the following result: all classical-quantum $2\times
N$ states  are necessarily SPPT. Hence, we provide a natural witness
for a quantum discord: if a $2\times N$ state is not SPPT it must
contain nonclassical correlations measured by quantum discord. It is
an analog of the celebrated Peres-Horodecki criterion: if a state is
not PPT, then it must be entangled.


\section{Quantum discord}


Let us briefly recall the definition of quantum discord
\cite{Zurek,Vedral}. Consider a density operator in $\mathcal{H}_A
\ot \mathcal{H}_B$ and let
\begin{equation}\label{}
    \mathcal{I}(\rho) = S(\rho_A) + S(\rho_B) - S(\rho) \ ,
\end{equation}
denote the quantum mutual information of a state $\rho$, where
$\rho_{A}$ $(\rho_B)$ is a reduced density matrix in $\mathcal{H}_A$
$(\mathcal{H}_B)$ and $S(\sigma)= - {\rm tr}(\rho \log\rho)$ stands
for the von Neumann entropy of the density operator $\sigma$. Note,
that mutual information may be rewritten as follows
\begin{equation}\label{}
    \mathcal{I}(\rho) = S(\rho_B)  - S(\rho|\rho_A) \ ,
\end{equation}
where $S(\rho|\rho_A)= S(\rho) - S(\rho_A)$ denotes quantum
conditional entropy. An alternative way to compute the conditional
entropy goes as follows: one introduces a measurement on $A$ part
defined by the collection of one-dimensional projectors $\{\Pi_k\}$
in $\mathcal{H}_A$ satisfying $\Pi_1 + \Pi_2 + \ldots =
\mathbb{I}_A$. The label `$k$' distinguishes different outcomes of
this measurement. The state after the measurement when the outcome
corresponding to $\Pi_k$ has been detected is given by
\begin{equation}\label{}
    \rho_{B|k} = \frac{1}{p_k} (\Pi_k \ot \mathbb{I}_B)\rho (\Pi_k \ot
    \mathbb{I}_B)\ ,
\end{equation}
where $p_k = {\rm tr}[\rho_{B|k} (\Pi_k\ot \mathbb{I}_B)]$. Hence,
$\rho_{B|k}$ defines an outcome of the local measurement conditioned
on the measurement outcome labeled by `$k$'. The entropies
$S(\rho_{B|k})$ weighted by probabilities $p_k$ yield to the
conditional entropy of part $B$ given the complete measurement
$\{\Pi_k\}$ on the part $A$
\begin{equation}\label{}
    S(\rho|\{\Pi_k\}) = \sum_k p_k S(\rho_{B|k})\ .
\end{equation}
Finally, let
\begin{equation}\label{}
    \mathcal{I}(\rho|\{\Pi_k\}) = S(\rho_B) - S(\rho|\{\Pi_k\}) \ ,
\end{equation}
be the corresponding measurement induced mutual information. The
quantity
\begin{equation}\label{C-sup}
    \mathcal{C}_{A}(\rho) = \sup_{\{\Pi_k\}} \mathcal{I}(\rho|\{\Pi_k\})\ ,
\end{equation}
is interpreted \cite{Zurek,Vedral} as a measure of  classical
correlations. Now, these two quantities -- $\mathcal{I}(\rho)$ and
$\mathcal{C}_A(\rho)$ -- may differ and the difference
\begin{equation}\label{}
    \mathcal{D}_{A}(\rho) =  \mathcal{I}(\rho) - \mathcal{C}_A(\rho)
\end{equation}
is called a quantum discord. For the definition of others
discord-like quantities see \cite{Terno,Wu}. Evidently, the above
definition is not symmetric with respect to parties $A$ and $B$.
However, one can easily swap the role of $A$ and $B$ to get
\begin{equation}\label{}
    \mathcal{D}_{B}(\rho) =  \mathcal{I}(\rho) - \mathcal{C}_B(\rho)
    \ ,
\end{equation}
where
\begin{equation}\label{C-sup}
    \mathcal{C}_{B}(\rho) = \sup_{\{\widetilde{\Pi}_\alpha\}} \mathcal{I}(\rho|\{\widetilde{\Pi}_\alpha\})\ ,
\end{equation}
and $\widetilde{\Pi}_\alpha$ is a collection of one-dimensional
projectors in $\mathcal{H}_B$ satisfying $\widetilde{\Pi}_1 +
\widetilde{\Pi}_2 + \ldots = \mathbb{I}_B$. For a general mixed
state $\mathcal{D}_A(\rho) \neq \mathcal{D}_B(\rho)$. However, it
turns out that $\mathcal{D}_A(\rho),\, \mathcal{D}_B(\rho) \geq 0$.
Moreover, on pure states, quantum discord coincides with the von
Neumann entropy of entanglement $S(\rho_A) = S(\rho_B)$. States with
zero quantum discord -- so called classical-quantum states --
represent essentially a classical probability distribution $p_k$
embedded in a quantum system. One shows that $\mathcal{D}_A(\rho)=0$
if and only if there exists an orthonormal basis $|k\>$ in
$\mathcal{H}_A$ such that
\begin{equation}\label{Q=0}
    \rho = \sum_k p_k\, |k\>\<k| \ot \rho^{(B)}_k \ ,
\end{equation}
where $\rho^{(B)}_k$ are density matrices in $\mathcal{H}_B$.
Similarly, $\mathcal{D}_B(\rho)=0$ if and only if there exists an
orthonormal basis $|\alpha\>$ in $\mathcal{H}_B$ such that
\begin{equation}\label{Q=0-B}
    \rho = \sum_\alpha q_\alpha\, \rho^{(A)}_\alpha \ot |\alpha\>\<\alpha| \ ,
\end{equation}
where $\rho^{(A)}_\alpha$ are density matrices in $\mathcal{H}_A$.
It is clear that if $\mathcal{D}_A(\rho)=\mathcal{D}_B(\rho)=0$,
then $\rho$ is diagonal in the product basis $|k\> \ot |\alpha\>$
and hence
\begin{equation}\label{Q=0-B}
    \rho = \sum_{k,\alpha} \lambda_{k\alpha}\, |k\>\<k| \ot |\alpha\>\<\alpha| \ ,
\end{equation}
is fully encoded by the classical joint probability distribution
$\lambda_{k\alpha}$. 

 In
this paper we consider only $\mathcal{D}_A$. Note, that $\Pi_k =
|k\>\<k|$ defines a measurement which is optimal for (\ref{C-sup}).
Hence, $\mathcal{D}_A(\rho)=0$ if
\begin{equation}\label{}
 \rho = \sum_k (\Pi_k \ot \mathbb{I}_B)\rho (\Pi_k \ot
    \mathbb{I}_B)\ .
\end{equation}
States with a positive quantum discord do contain nonclassical
correlations  even if they are separable. Hence nonvanishing quantum
discord indicates a kind of  quantumness encoded in a separable
mixed state. Actually, there is a simple necessary criterion for
zero quantum discord \cite{Acin}: if $\mathcal{D}_A(\rho)=0$, then
\begin{equation}\label{Ac}
    [\rho,\rho_A \ot \mathbb{I}_B]= 0 \ .
\end{equation}
Hence, if $\rho$ does not commute with $\rho_A \ot \mathbb{I}_B$ its
quantum discord is strictly positive and, hence, $\rho$ is
nonclassically correlated.   This quantumness may we associated for
example to the impossibility of local broadcasting \cite{B1,B2}. For
the recent discussion of zero discord states see
\cite{null-1,null-2}.



\section{Main result}

Any state of a bipartite system living in $\mathbb{C}^2 \ot
\mathbb{C}^N$ may be considered as a block $2 \times 2$ matrix with
$N \times N$ blocks. Positivity of $\rho$ implies that $\rho =
\mathbf{X}^\dagger \mathbf{X}$ for some $2\times 2$ upper triangular
block matrix $\bf X$ (due to the well known Cholesky decomposition)
\begin{equation}\label{X}
\mathbf{X} = \left( \begin{array}{c|c} X_1 & SX_1  \\ \hline
  0 & X_2  \end{array} \right)\ ,
\end{equation}
with arbitrary $N \times N$ matrices $X_1,X_2$ and $S$.  One finds
\begin{equation}\label{XX-2}
    \rho = \mathbf{X}^\dagger \mathbf{X} =  \left( \begin{array}{c|c} X_1^\dagger  X_1 & X_1^\dagger  S X_1  \\
\hline  X_1^\dagger  S^\dagger  X_1 & X_1^\dagger  S^\dagger  S X_1
+ X_2^\dagger  X_2
\end{array} \right) \ ,
\end{equation}
and  for its partial transposition
\begin{equation}\label{T-rho}
    \rho^{{\rm T}_A} =
\left( \begin{array}{c|c} X_1^\dagger  X_1 & X_1^\dagger  S^\dagger  X_1  \\
\hline  X_1^\dagger S X_1 & X_1^\dagger  S^\dagger  S X_1 +
X_2^\dagger X_2
\end{array} \right) \ .
\end{equation}
Note, that there is a gauge freedom in choosing $X_1,X_2$ and $S$:
one may perform the following transformation
\begin{equation*}\label{}
    X_1 \longrightarrow G_1X_1\ , \ \ \ X_2 \longrightarrow G_2X_2\
    ,\ \ \ S \longrightarrow G_1SG_1^{-1}\ ,
\end{equation*}
with $G_1,G_2 \in U(N)$, leaving the formula for $\rho$ invariant.
In particular, one can always take $X_1$ to be semipositive
definite. Clearly, $\rho$ is PPT iff there exists $\bf Y$ such that
$\rho^{{\rm T}_A} = {\bf Y}^\dagger {\bf Y}$. The choice of $\bf Y$
(if it exists) is highly nonunique. Note, however, that there is a
`canonical' candidate for $2N \times 2N$ matrix $\bf Y$ defined by
(\ref{X}) with $S$ replaced by $S^\dagger$, that is
\begin{equation}\label{Y}
{\bf Y} = \left( \begin{array}{c|c} X_1 & S^\dagger X_1  \\
\hline  0 & X_2  \end{array} \right) \ ,
\end{equation}
and hence
\begin{equation}\label{YY}
    {\bf Y}^\dagger {\bf Y} =
\left( \begin{array}{c|c} X_1^\dagger  X_1 & X_1^\dagger  S^\dagger  X_1  \\
\hline  X_1^\dagger   S X_1 & X_1^\dagger  S S^\dagger  X_1 +
X_2^\dagger X_2
\end{array} \right) \ .
\end{equation}
Let us observe that if $S$ is normal, that is,
\begin{equation}\label{SS}
 S^\dagger  S  =  S S^\dagger \ ,
\end{equation}
then $\rho^{T_A} = {\bf Y}^\dagger {\bf Y}$ and hence $\rho$ is PPT.
We call such PPT states --- SPPT states \cite{SPPT}. Note, that
condition (\ref{SS}) is gauge invariant, that is, if $S$ satisfies
(\ref{SS}) so does $S' = G_1 S G_1^{-1}$. For a generalization of
SPPT for $M \times N$ systems cf. \cite{SPPT}.



The main result of our paper consists in the following

\begin{theorem}
If $\mathcal{D}_A(\rho)=0$, then $\rho$ is SPPT.
\end{theorem}
To prove it let us observe that $\mathcal{D}_A(\rho)=0$ implies that
there exists a basis $\{f_1,f_2\}$ in $\mathbb{C}^2$ such that
\begin{equation}\label{D=0}
    \rho = \sum_{i=1}^2 |f_i\>\<f_i| \ot \sigma_i \ ,
\end{equation}
where $\sigma_i \geq 0$ and ${\rm Tr}(\sigma_1 + \sigma_2)=1$. Let
$U$ be a unitary in $\mathbb{C}^2$ and let $|f_i\> = U|e_i\>$. The
block structure of (\ref{D=0}) in the canonical computational basis
$\{e_1,e_2\}$ reads as follows
\begin{equation}\label{}
\rho = \left( \begin{array}{c|c} \rho_{11} & \rho_{12}  \\ \hline
  \rho_{21} & \rho_{22}  \end{array} \right)\ ,
\end{equation}
where
\begin{eqnarray}
  \rho_{11} &=& |U_{11}|^2 \sigma_1 + |U_{12}|^2\sigma_2\ ,\nonumber  \\
  \rho_{22} &=& |U_{21}|^2 \sigma_1 + |U_{22}|^2\sigma_2\ , \\
  \rho_{12} &=& U_{11}\overline{U}_{21}\,\sigma_1 + U_{12}\overline{U}_{22}\,\sigma_2\
  , \nonumber
\end{eqnarray}
and $\rho_{21} = \rho_{12}^\dagger$. One has therefore
\begin{eqnarray}
  X_1^\dagger X_1 = |U_{11}|^2 \sigma_1 + |U_{12}|^2\sigma_2\ ,
\end{eqnarray}
and hence one may take
\begin{equation}\label{X1}
    X_1 = \left( |U_{11}|^2 \sigma_1 + |U_{12}|^2\sigma_2
    \right)^{1/2}\ .
\end{equation}
Clearly, $X_1$ is hermitian and semipositive definite $X_1 \geq 0$.
Assume now that $X_1$ is full rank $N \times N$ matrix, that is,
$X_1$ is strictly positive. Then
\begin{eqnarray}
X_1^\dagger  S  X_1   = U_{11}\overline{U}_{21} \sigma_1 +
U_{12}\overline{U}_{22} \sigma_2\ ,
\end{eqnarray}
gives rise to the following formula for $S$
\begin{equation}\label{S}
    S = X_1^{-1} \left( \, U_{11}\overline{U}_{21} \sigma_1 +
U_{12}\overline{U}_{22} \sigma_2     \right) X_1^{-1}\ .
\end{equation}
If $X_1$ is not strictly positive we may take the generalized
inverse (so called Moore-Penrose pseudoinverse).  Finally, taking
into account
\begin{equation}\label{UU}
    U_{11} \overline{U}_{21} + U_{12} \overline{U}_{22} = 0 \ ,
\end{equation}
one obtains
\begin{equation}\label{}
    S = U_{11} \overline{U}_{21}\, X_1^{-1} \left( \sigma_1 - \sigma_2
    \right) X_1^{-1}\ .
\end{equation}
Note, that since $X_1^{-1} \left( \sigma_1 - \sigma_2 \right)
X_1^{-1}$ is hermitian, $S$ is normal which ends the proof.

\begin{corollary} If a PPT state $\rho$ in $\mathbb{C}^2 \ot
\mathbb{C}^N$ is not SPPT, then the quantum discord of $\rho$ does
not vanish.
\end{corollary}
\begin{remark} {\em It turns out that this result does not hold for general $M
\times N$ systems with $M >2$. Consider for example $M=3$. One can
easily introduce $3\times N$ SPPT states as follows \cite{SPPT}: let
$\rho = \mathbf{X}^\dagger \mathbf{X}$ for some $3\times 3$ upper
triangular block matrix $\bf X$
\begin{equation}\label{X-3}
\mathbf{X} = \left( \begin{array}{c|c|c} X_1 & S_{12} X_1 & S_{13} X_1  \\
\hline   0 & X_2 & S_{23} X_2 \\ \hline 0 & 0 & X_3  \end{array}
\right)\ ,
\end{equation}
with arbitrary $N \times N$ matrices $X_1,X_2,X_3$ and
$S_{12},S_{13},S_{23}$. Now, $\rho$ is SPPT \cite{SPPT} if all three
matrices $S_{kl}$ are normal and
\begin{equation}\label{}
    S_{12} S_{13}^\dagger = S_{13}^\dagger S_{12} \ .
\end{equation}
One easily finds for the block structure
\begin{equation}\label{}
\rho = \left( \begin{array}{c|c|c} \rho_{11} & \rho_{12} & \rho_{13} \\
\hline   \rho_{21} & \rho_{22} & \rho_{23} \\ \hline \rho_{31} &
\rho_{32} & \rho_{33}   \end{array} \right)\ ,
\end{equation}
where
\begin{eqnarray} \label{rho-kl}
  \rho_{11} &=& X_1^\dagger X_1 \ , \nonumber \\
  \rho_{1k} &=& X_1^\dagger S_{1k} X_1 \ , \ \ \ \ \ \ \  k=2,3 \ ,\nonumber \\
  \rho_{22}  &=& X_1^\dagger S_{12}^\dagger S_{12} X_1 + X_2^\dagger  X_2\ , \\
  \rho_{23}  &=& X_1^\dagger S_{12}^\dagger S_{13} X_1 + X_2^\dagger S_{23}  X_2\ , \nonumber \\
  \rho_{33} &=& X_1^\dagger S_{13}^\dagger S_{13} X_1 + X_2^\dagger S_{23}^\dagger S_{23}  X_2 + X_3^\dagger X_3\
  . \nonumber
\end{eqnarray}
Now, $\mathcal{D}_A(\rho)=0$ if there exists an orthonormal basis
$\{f_1,f_2,f_3\}$ in $\mathbb{C}^3$ such that
\begin{equation}\label{}
    \rho = \sum_{k=1}^3 |f_k\>\<f_k| \ot \sigma_k \ ,
\end{equation}
with $\sigma_k \geq 0$ and ${\rm Tr}(\sigma_1 + \sigma_2 +
\sigma_3)=1$. Let $U$ be a unitary operator defined by $|e_k\> =
U|f_k\>$. One has
\begin{equation}\label{}
    \rho_{kl}= \sum_{m=1}^3 U_{km} \overline{U}_{lm} \sigma_m \ .
\end{equation}
Therefore formula (\ref{rho-kl}) gives for $\rho_{11}$
\begin{eqnarray}
  X_1^\dagger X_1 = |U_{11}|^2 \sigma_1 + |U_{12}|^2\sigma_2 + |U_{13}|^2\sigma_3\ ,
\end{eqnarray}
and hence one may take
\begin{equation}\label{X1}
    X_1 = \left( |U_{11}|^2 \sigma_1 + |U_{12}|^2\sigma_2
    +|U_{13}|^2\sigma_3
    \right)^{1/2}\ .
\end{equation}
Clearly, $X_1$ is hermitian and semipositive definite $X_1 \geq 0$.
Assume now that $X_1$ is strictly positive. Then formula
(\ref{rho-kl}) gives the following formula for $S_{12}$
\begin{equation*}\label{}
    S_{12} = X_1^{-1} \left( \, U_{11}\overline{U}_{21} \sigma_1 +
U_{12}\overline{U}_{22} \sigma_2 + U_{13}\overline{U}_{23} \sigma_3
    \right) X_1^{-1}\ .
\end{equation*}
Now, contrary to $S$ defined by (\ref{S}) $S_{12}$ needs not be
normal. Using
\begin{equation}\label{}
    U_{11} \overline{U}_{21} + U_{12} \overline{U}_{22} + U_{13} \overline{U}_{23}= 0 \ ,
\end{equation}
one obtains
\begin{equation}\label{}
    S_{12} = \lambda_1 H_1 + \lambda_2 H_2 \ ,
\end{equation}
where the complex numbers $\lambda_k$ are defined by
\begin{equation*}\label{}
    \lambda_1 = U_{12} \overline{U}_{32}\ , \ \ \ \lambda_2 = U_{13}
    \overline{U}_{33}\ ,
\end{equation*}
and Hermitian operators $H_1$ and $H_2$ reads as follows
\begin{equation*}\label{}
    H_1 = X_1^{-1}(\sigma_2-\sigma_1) X_1^{-1} \ , \ \ \  H_2 = X_1^{-1}(\sigma_3-\sigma_1)
    X_1^{-1}\ .
\end{equation*}
Hence
\begin{equation}\label{}
    [S_{12},S_{12}^\dagger] = (\lambda_1\overline{\lambda}_2 - \lambda_2\overline{\lambda}_1)\,
    [H_1,H_2]\ ,
\end{equation}
which shows that in general the commutator $[S_{12},S_{12}^\dagger]$
does not vanish and hence $S_{12}$ is not normal.

 }
\end{remark}



\section{Example -- $X$-states}

To illustrate our analysis let us consider so called $X$-state of
two qubits \cite{Lu,X-recent}
\begin{equation}\label{X-state}
\rho = \left( \begin{array}{cc|cc} a_{11} & \cdot & \cdot & a_{12}  \\
                                    \cdot & b_{11}& b_{12}& \cdot \\ \hline
                                    \cdot & b_{21}& b_{22}& \cdot \\
                                    a_{21} & \cdot & \cdot & a_{22} \end{array} \right)\ ,
\end{equation}
where to make the picture more transparent we replaced all zeros by
dots. The matrices
\begin{equation}\label{}
a = \left( \begin{array}{cc} a_{11} & a_{12}  \\
  a_{21} & a_{22}  \end{array} \right)\ , \ \ \ b = \left( \begin{array}{cc} b_{11} & b_{12}  \\
  b_{21} & b_{22}  \end{array} \right)\ ,
\end{equation}
satisfy: $a \geq 0$, $b \geq 0$ and ${\rm Tr}(a+b)=1$. Clearly, if
$a_{12}=b_{12}=0$, a state is separable with
$\mathcal{D}_A(\rho)=0$. If only one off-diagonal element $a_{12}$
or $b_{12}$   is different from zero, then $\rho$ is necessarily
entangled being NPT. Hence, let us assume that both $a_{12} \neq 0$
and $b_{12}\neq 0$. Note that partially transposed state $\rho^{{\rm
T}_A}$ has again an $X$-structure with matrices $a$ and $b$ replaced
by
\begin{equation}\label{}
\widetilde{a} = \left( \begin{array}{cc} a_{11} & b_{21}  \\
  b_{12} & a_{22}  \end{array} \right)\ , \ \ \ \widetilde{b} = \left( \begin{array}{cc} b_{11} & a_{21}  \\
  a_{12} & b_{22}  \end{array} \right)\ .
\end{equation}
Hence, $X$-state $\rho$ is PPT iff $\widetilde{a}\geq 0$ and
$\widetilde{b} \geq 0$. Now, positivity of $\rho$ is equivalent to
\begin{equation}\label{P1}
    a_{11}a_{22} \geq |a_{12}|^2\ , \ \ \ b_{11}b_{22} \geq |b_{12}|^2\
    .
\end{equation}
A state is PPT if additionally one has
\begin{equation}\label{P2}
    a_{11}a_{22} \geq |b_{12}|^2\ , \ \ \ b_{11}b_{22} \geq |a_{12}|^2\
    .
\end{equation}
One shows \cite{SPPT} that a state is SPPT iff
\begin{equation}\label{P3}
    |a_{12}| = |b_{12}|\ .
\end{equation}
Clearly, (\ref{P3}) implies (\ref{P2}). Finally, following our
analysis it is easy to show that $\rho$ has vanishing discord iff it
satisfies (\ref{P3}) and
\begin{equation}\label{}
    a_{11} = b_{22}\ , \ \ \ a_{22}=b_{11} \ .
\end{equation}
Hence, $\mathcal{D}_A(\rho)=0$ if and only if matrices $||a_{ij}||$
and $||b_{ij}||$ are unitarily equivalent $b = V a V^\dagger$ with
\begin{equation}\label{}
    V = \left( \begin{array}{cc} 0 & e^{i\mu}   \\
  e^{i\nu} & 0  \end{array} \right)\ , \ \ \ \ \mu,\nu \in
  \mathbb{R}\ .
\end{equation}
 Therefore, one has the following chain of proper inclusions
\[  \{\mathcal{D}_A=0\} \subset {\rm SPPT } \subset {\rm PPT}\ . \]

In particular if $\rho$ is Bell diagonal, i.e.
\begin{eqnarray*}
  a_{11} &=& a_{22} =  p_1+p_2 \ , \\
  a_{12} &=& p_1-p_2 \ ,\\
  b_{11} &=& b_{22} = p_3+p_4\ , \\
  b_{12} &=& p_3-p_4\ ,
\end{eqnarray*}
where $p_k \geq 0$ and $p_1+p_2+p_3+p_4=1$, then $\rho$ is SPPT if
$\,|p_1-p_2| = |p_3-p_4|\,$.  Moreover, $\mathcal{D}_A(\rho)=0$ if
and only if 1) $p_1=p_3$ and $p_2=p_4$, or 2) $p_1=p_4$ and
$p_2=p_3$. Hence, discord zero Bell diagonal state of 2 qubits has
the following form
\begin{equation}\label{Bell}
\rho = \frac 14 \left( \begin{array}{cc|cc} 1 & \cdot & \cdot & q  \\
                                    \cdot & 1 & \pm q & \cdot \\ \hline
                                    \cdot & \pm q & 1& \cdot \\
                                    q & \cdot & \cdot & 1 \end{array} \right)\ ,
\end{equation}
where $-1 \leq q \leq 1$. This results do agree with  the analysis
of $X$-state performed in \cite{Lu} and recently in \cite{X-recent}.
Note, that the above formula defines 1-dimensional subset in the
3-dimensional set of Bell-diagonal states. Let us observe that for
Bell diagonal states $\rho_A = \rho_B = \mathbb{I}_2/2$ and hence
the condition (\ref{Ac})
is satisfied for all Bell diagonal states. It shows that a necessary
criterion of zero quantum discord \cite{Acin} cannot detect discord
within this class. Note, that (\ref{Bell}) implies that any convex
combination $\,\frac 12 (P_1 + P_2) \,$ of arbitrary two Bell
projectors $P_1$ and $P_2$ has vanishing discord.

\section{Conclusions}

We provided a simple witness for a nonclassical correlations
measured by a quantum discord in $2\times N$ systems. We stress that
our result is not true for $M \times N$ system with $M>2$.  Note the
similarity with Peres-Horodecki criterion. Being PPT is equivalent
to separability only for $2\times 2$ and $2\times 3$ systems.
It would be interesting to look for the condition
$\mathcal{D}_A(\rho)=0$ for the generalization of $X$-states for $d
\times d$ system. Such states were constructed in \cite{CIRC} (we
called them circulant states, see also \cite{C1}). In particular
they provide generalization of Bell diagonal states of two qudits.


\noindent {\it Acknowledgments}. This work was partially supported
by the Polish Ministry of Science and Higher Education Grant No
3004/B/H03/2007/33.

\end{document}